\documentstyle[aps,multicol,epsfig]{revtex}

\def\d{{\partial}}
\def\s{{\sigma}}
\def\e{{\epsilon}}
\def\k{{ {\bf k} }}

\def\q{{ {\bf q} }}

\def\w{{\omega}}
\def\a{{\alpha}}

\begin{document}
\draft

\def\runtitle{
Nernst coefficient and Magnetoresistance
in High-$T_{\rm c}$ Superconductors
}
\def\runauthor
 {Hiroshi {\sc Kontani}}

\title{
Nernst coefficient and Magnetoresistance
in High-$T_{\rm c}$ Superconductors :\\
the Role of Superconducting Fluctuations
}

\author{
Hiroshi {\sc Kontani}
}

\address{
Department of Physics, Saitama University,
255 Shimo-Okubo, Saitama-city, 338-8570, Japan.
}

\date{\today}

\maketitle      

\begin{abstract}
In hole-doped high-$T_{\rm c}$ cuprates,
the Nernst coefficient ($\nu$) as well as the magnetoresistance 
($\Delta\rho/\rho$)
increase drastically below the pseudo-gap temperature, $T^\ast$.
This unexpected result attracts much attention in that 
it reflects the fundamental feature of the electronic
state in the pseudo-gap region, which has been a
central issue on high-$T_{\rm c}$ cuprates.
In this letter, we study these transport phenomena
in terms of the fluctuation-exchange (FLEX)+T-matrix approximation.
In this present theory,
the $d$-wave superconducting (SC) fluctuations,
which are mediated by antiferromagnetic (AF) correlations,
become dominant below $T^\ast$.
We focus on the role of the vertex corrections 
both for the charge current and the heat one,
which are indispensable to keep the conservation laws.
As a result,
the mysterious behaviors of $\nu$ and $\Delta\rho/\rho$, 
which are key phenomena in the pseudo-gap region,
are naturally explained as the reflection of the
enhancement of the SC fluctuation,
without assuming thermally excited vortices.
The present result suggests that
the pseudo-gap region in high-$T_c$ cuprates
is well described in terms of the Fermi liquid
with AF and SC fluctuations.

\end{abstract}



\begin{multicols}{2}

In high-Tc cuprates,
various transport phenomena in the normal state
show striking non-Fermi liquid (NFL)-like behaviors,
such as resistivity ($\rho$)
 \cite{rho-exp}, 
Hall coefficient ($R_{\rm H}$)
 \cite{Hall-exp},
magnetoresistance (MR, $\Delta\rho/\rho$)
 \cite{MR-exp,MR-exp2,Ando-MR}, 
thermoelectric power (TEP, $S$)
 \cite{S-exp}
and Nernst coefficient ($\nu$)
 \cite{N-exp,N-exp2}.
Because
it is impossible to understand these NFL-like phenomena
in the framework of the relaxation time approximation (RTA),
one may consider that the Landau-Fermi liquid picture,
--- concept of the quasiparticle ---, is totally
violated in high-Tc cuprates.
However, before making a conclusion on this issue,
one has to study the transport coefficients
beyond the RTA, 
e.g., by the ``conserving'' approximation as Baym and Kadanoff
 \cite{Baym}.

In these years,
we have studied $R_{\rm H}$
\cite{Hall},
$\Delta\rho/\rho$
\cite{MR,MR-HTSC}
and $S$
\cite{S-HTSC}
in high-Tc cuprates above the pseudo-gap temperature ($T^\ast$)
in terms of the conserving approximation:
Owing to the vertex corrections (VC's) for currents,
the total charge current with VC's, ${\vec J}_\k$,
is no more perpendicular to the Fermi surface
in the presence of strong antiferromagnetic (AF) fluctuations.
This important mechanism of the VC, i.e., 
the ``back-flow'' in the hydrodynamic regime,
had been overlooked for years.
 \cite{Hall}.
As a result, we succeeded to reproduce
various NFL-like transport phenomena 
qualitatively as for $T>T^\ast$.
 \cite{SCR}.

In this letter,
we study transport phenomena below $T^\ast$.
In the present stage,
superconducting (SC) fluctuation is
one of the promising origins of the pseudo-gap phenomena 
 \cite{Levin,Koikegami,Yanase,Kobayashi}.
Based on this opinion, 
we combine the fluctuation-exchange (FLEX) approximation
for describing the AF fluctuations
 \cite{Bickers}
with the $T$-matrix approximation
for describing the SC fluctuations
(i.e., FLEX+$T$-matrix approximation)
  \cite{Koikegami,Yanase,Kobayashi}.
Next, we study the Nernst coefficient,
$\nu\equiv S_{yx}/B =-E_y/B \d_x T$,
which is an off-diagonal TEP under the magnetic field
${\bf B} \parallel {\bf z}$.
According to the linear response theory,
\begin{eqnarray}
\nu &=& \left[ \ {\a_{xy}}/{\s}-S\tan\theta_{\rm H}\ \right]/B,
\end{eqnarray}
where 
$\a_{xy}$ is the off-diagonal Peltier conductivity
($J_x = \a_{xy}(-\d_y T)$),
and $\tan\theta_{\rm H}\equiv \s_{xy}/\s$.
In high-$T_{\rm c}$ cuprates,
$\nu$ increases divergently below $T^\ast$,
which is sometimes interpreted 
as the evidence for the spontaneous vortex-like excitations
in the pseudo-gap region
 \cite{N-exp,N-exp2}.
[In the the mixed state,
$\nu$ takes very large value 
in a clean 2D system,
reflecting the high-mobility of a vortex.]
Contrary to such an exotic scenario,
in the present work,
we study $\nu$ due to the heat current
carried by the quasiparticle motion.
By considering the VC's due to strong AF and SC fluctuations,
we obtain the results which are highly consistent with experiments.
Thus, the nature of the pseudo-gap region
is well described as the 
Fermi liquid with the strong AF+SC fluctuations.

In the self-consistent FLEX+$T$-matrix approximation,
the full Green function and the self-energy are given by
\begin{eqnarray}
& &G_\k(\e_n) = (i\e_n+\mu-\e_\k^0-\Sigma_\k(\e_n))^{-1},
 \label{eqn:Green} \\
& &\Sigma_\k(\e_n) =
 \Sigma_\k^{\rm FLEX}(\e_n) + \Sigma_\k^{\rm SCF}(\e_n),
 \label{eqn:Self} 
\end{eqnarray}
where $\e_\k^0$ is the tight binding dispersion
and $\e_n$ is a fermion Matsubara frequency.
$\Sigma^{\rm FLEX}$ is given by the diagrams
for the FLEX approximation, and
$\Sigma^{\rm SCF}$ is given by the $T$-matrix approximation
 \cite{Koikegami,Yanase,Kobayashi}.
This is a kind of the one-loop approximation with respect to 
the AF and SC fluctuations.
In the pseudo-gap region,
$\Sigma^{\rm SCF}$ is approximately given by
 \cite{Levin}
\begin{eqnarray}
\Sigma_\k^{\rm SCF}(\e_n) &=& -\Delta^2_{\rm pg}\psi_\k^2
 G_\k(-\e_n) ,
 \label{eqn:S-SCF} \\
\Delta_{\rm pg}^2&=& T{\sum_\q}' t_{\rm pg}(\q,\w=0) ,
 \label{eqn:Dqp}
\end{eqnarray}
where $\psi_\k= \cos k_x -\cos k_y$ and
$\sum'_\q \equiv \sum_{\q}\theta(a-|\q|)$.
The factor $a\ll1$ is introduced
not to overestimate the effect of SC fluctuations.
We put $a=0.03\pi$ in the present calculation,
which would be smaller than the inverse of the 
SC coherent length.
The result is not sensitive to its value, and
this approximation becomes reasonable as $T\rightarrow T_c$,
which is favorite for the purpose of our study.
$t_{\rm pg}(\q,\w)$ is the $T$-matrix for $d_{x^2\mbox{-}y^2}$-wave channel,
which is mediated by strong AF fluctuations.

\begin{figure}
\begin{center}
\epsfig{file=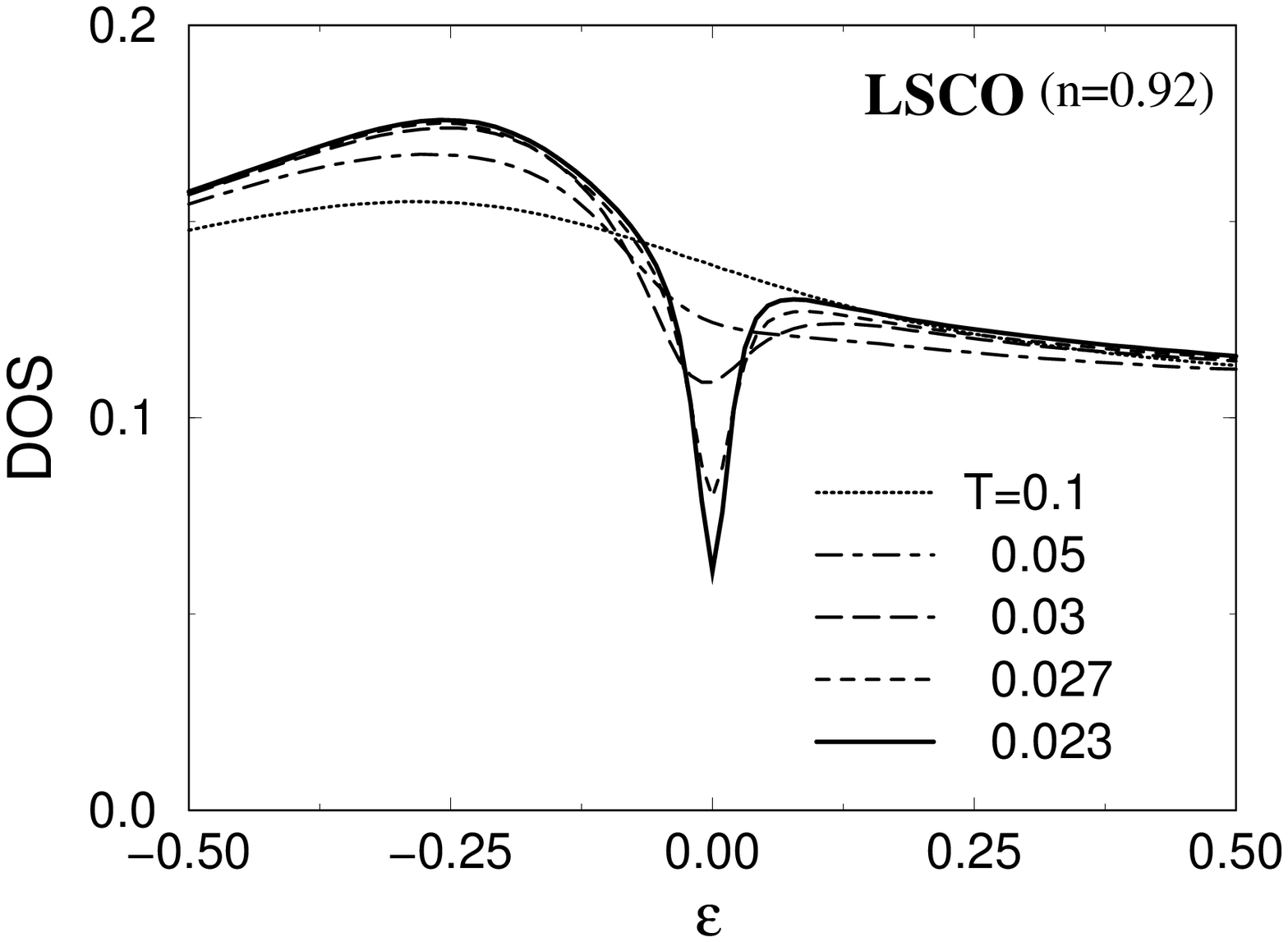,width=5.4cm}
\epsfig{file=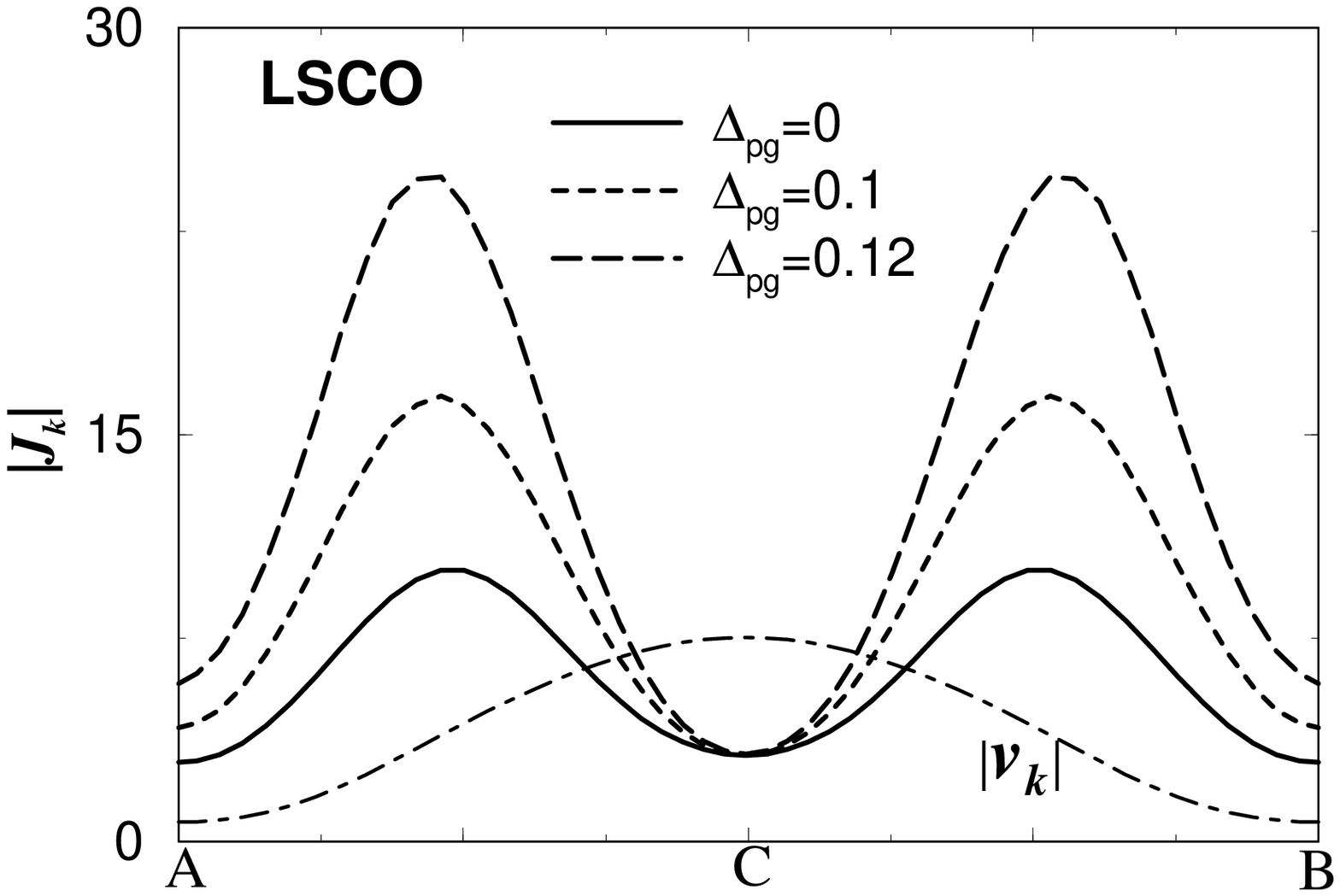,width=5.4cm}
\epsfig{file=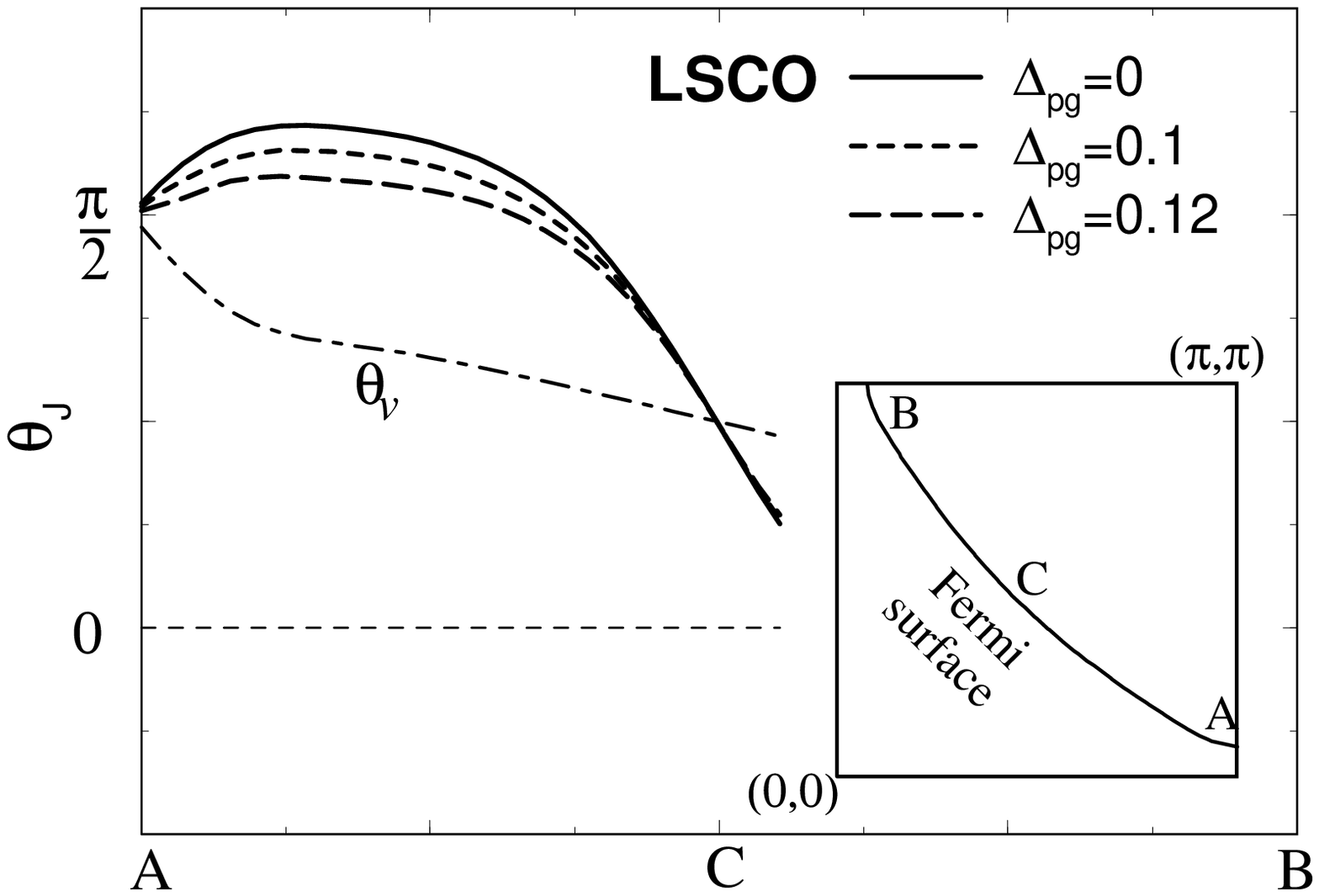,width=5.4cm}
\end{center}
\caption{
(i) The DOS obtained by the self-consistent FLEX+T-matrix 
approximation.
$T=0.1$ corresponds to $300\sim400$K.
(ii,iii) $|{\vec J}_\k|$ and $\theta_\k^J$
obtained by the B-S equation (\ref{eqn:J-SF-SC}) 
for various $\Delta_{\rm pg}$.
}
  \label{fig:J}
\end{figure}

To derive $t_{\rm pg}$,
we solve the Eliashberg equation:
\begin{eqnarray}
\lambda \phi_\k(\e_n)= -T\sum_{\k',\e_m}
 V_{\k-\k'}^{\rm FLEX}(\e_{n}-\e_m)
 |G_{\k'}(\e_{m})|^2 \phi_{\k'}(\e_{m}),
 \nonumber
\end{eqnarray} 
where $V_\k^{\rm FLEX}$ is the effective interaction
for a singlet pair within the FLEX approximation, 
which is given by eq.(9) of ref.\cite{Hall}
using the Green function in eq.(\ref{eqn:Green}).
$\lambda$ is the eigenvalue of the equation
which exceeds 1 below $T_c$.
Hereafter, the eigenfunction $\phi$ is normalized as 
$\sum_{\k,l}\phi_\k^2(\e_l)=1$.
By using $\phi$,
we can approximate that 
\begin{eqnarray}
V_{\k-\k'}^{\rm FLEX}(\e-\e') \approx
  g\phi_\k(\e)\phi_{\k'}(\e'),
 \label{eqn:V-ap}
\end{eqnarray}
where
$g \equiv \sum_{\e,\e'}\sum_{\k,\k'}
 V_{\k-\k'}^{\rm FLEX}(\e-\e')\phi_\k(\e)\phi_{\k'}(\e')$.
Then, the  $T$-matrix is given by
\begin{eqnarray}
t_{\rm pg}(q,\w_l)&=& g/(1+g{\bar \chi}_\q(\w_l)),
 \\
{\bar \chi}_\q(\w_l)
 &=& T\sum_{\k,\e_n}G_\k(\e_n)G_{\q-\k}(\w_l-\e_n)
 \phi_\k^2(\e_n) ,
 \label{eqn:t-mat}
\end{eqnarray}
where $\w_l$ is a Matsubara frequency for boson.
It is easy to see that
$\lambda = -g{\bar \chi}_{\q={\bf 0}}(0)$
according to eq.(\ref{eqn:V-ap}).
In the FLEX+$T$-matrix approximation,
we solve eqs.(\ref{eqn:Green})-(\ref{eqn:t-mat})
self-consistently.
In this approximation,
$\lambda<1$ is satisfied in 2D systems 
at finite temperatures
because a Kosterlitz-Thouless type 
transition is not taken into account.

In the present numerical study
for the Hubbard model,
we put $U=4.5$ and $(t,t',t'')=(-1,0.15,-0.05)$ for 
La$_{2-x}$Sr$_x$CuO$_4$ (LSCO).
where $t,t',t''$ are the nearest, the next-nearest and 
the third-nearest neighbor hoppings, respectively.
Figure \ref{fig:J}
shows the obtained 
density of states (DOS);
$\rho(\e)= \frac1{\pi}\sum_\k{\rm Im}G_\k(\e-i\delta)$.
We see that
a large pseudo-gap emerges below $T^\ast \sim 0.03$,
because SC fluctuations grow prominently below $T^\ast$
 \cite{Koikegami,Yanase,Kobayashi}.
In the present self-consistent calculation,
$\lambda = 0.988$ and $\Delta_{\rm pg}=0.147$ at T=0.02.


Next, we calculate the Nernst coefficient
by taking VC's into account.
Based on the linear response theory
for the thermoelectric transport phenomena
 \cite{Mahan},
we can derive the general expression for $\a_{xy}$
in correlated electron systems
by referring the derivation for $\s_{xy}$
by Kohno and Yamada  
 \cite{Kohno}.
The VC is uniquely given by 
the Ward identities associated with
the local charge and energy conservation laws
 \cite{Nernst-formula}.
The obtained expression,
which 
is exact with respect to $O(\gamma^{-2})$,
is given by
\begin{eqnarray}
\a_{xy} &=& B \cdot \frac{e^2}{T}\sum_\k\int \frac{d\e}{2\pi} 
 \left(-\frac{\d f}{\d\e}\right) |{\rm Im}G_\k(\e)||G_\k(\e)|^2 
 \nonumber \\
& &\times |{\vec v}_{\k}(\e)| \gamma_\k(\e) A_\k(\e),
 \label{eqn:axy} \\
A_\k(\e)&=& \left( {\vec Q}_{\k}(\e) 
 \times \frac{\d}{\d k_\parallel} 
 \left({\vec J}_{\k}(\e) / \gamma_\k(\e) \right)
 \right)_z,
 \\
{\vec Q}_{\k}(\e) \! &=& \! {\vec q}_{\k}(\e)
 + \sum_{\k'} \!\! \int \!\! \frac{d\e'}{4\pi i}
 {\cal T}_{\k\k'}(\e,\e')|G_{\k'}(\e')|^2 {\vec Q}_{\k'}(\e') ,
 \!\! \label{eqn:Q}\\
{\vec J}_{\k}(\e) \! &=& \! {\vec v}_{\k}(\e)
 + \sum_{\k'} \!\! \int \!\! \frac{d\e'}{4\pi i}
 {\cal T}_{\k\k'}(\e,\e')|G_{\k'}(\e')|^2 {\vec J}_{\k'}(\e') ,
 \!\! \label{eqn:J}
\end{eqnarray}
where $k_\parallel$ is the momentum along the Fermi surface,
${\vec v}_{\k}(\e)\equiv 
{\vec\nabla}(\e_\k^0+{\rm Re}\Sigma_\k(\e))$ and
${\vec q}_{\k}(\e)\equiv \e \cdot {\vec v}_{\k}(\e)$.
${\cal T}$ is the irreducible VC
introduced by Eliashberg as ${\cal T}_{22}^{(0)}$
 \cite{Eliashberg}.

Next, we calculate the total charge current ${\vec J}$ 
and the heat one ${\vec Q}$ numerically.
Here,
we take only the infinite series of the
Maki-Thompson (MT) type VC's 
due to the AF and SC fluctuations 
although the Ward identity gives other terms,
because they are expected to be most important
when the fluctuations are strong
 \cite{Hall,Yanase}.
In the present FLEX+$T$-matrix approximation,
the Bethe-Sapleter (B-S) equation (\ref{eqn:J})
is simply written as
\begin{eqnarray}
{\vec J}_{\k}(\e)&=& 
 \left({\gamma_\k(\e)/\gamma_\k^{\rm FLEX}(\e) }\right)
 \cdot \Bigl( {\vec v}_{\k x}(\e) 
 \nonumber \\
& & + \sum_{\k'}\int\frac{d\e'}{4\pi i}
 {\cal T}_{\k,\k'}^{\rm FLEX}(\e,\e')
 |G_{\k'}(\e')|^2 {\vec J}_{\k'}(\e')
 \Bigr) ,
 \label{eqn:J-SF-SC}
\end{eqnarray}
where ${\cal T}^{\rm FLEX}$
represents the MT-terms due to the AF fluctuations,
whose functional form is given by eq.(A9) of ref.
 \cite{Hall},
Note that the factor ${\gamma_\k/\gamma_\k^{\rm FLEX}}$
comes from the MT-terms due to the SC fluctuations,
which becomes unity if $\Delta_{\rm qp}=0$.
Performing the similar discussion in \S V of ref.
 \cite{Hall}
and using the relation
$\gamma_\k^{\rm FLEX}
= \sum_{\k'}\int\frac{d\e}{4\pi}
 {\cal T}_{\k,\k'}^{\rm FLEX}(0,\e){\rm Im}G_{\k'}^{\rm A}(\e)$,
the following approximate relation
is derived from eq.(\ref{eqn:J-SF-SC})
 \cite{future}:
\begin{eqnarray}
{\vec J}_\k \ \approx \ 
 ({\gamma_\k}/{\gamma_\k^{\rm FLEX}})
{\vec J}_\k^{\ [\Delta_{\rm qp}=0]} ,
 \label{eqn:J-ap}
\end{eqnarray}
where ${\vec J}_\k^{\ [\Delta_{\rm qp}=0]}$ 
is given by the FLEX approximation 
without $\Sigma_\k^{\rm SCF}$ and 
${\cal T}^{\rm SCF} = {\cal T}-{\cal T}^{\rm FLEX}$, 
whose behavior was analysed in ref. 
 \cite{Hall}
in detail.
 
\begin{figure}
\begin{center}
\epsfig{file=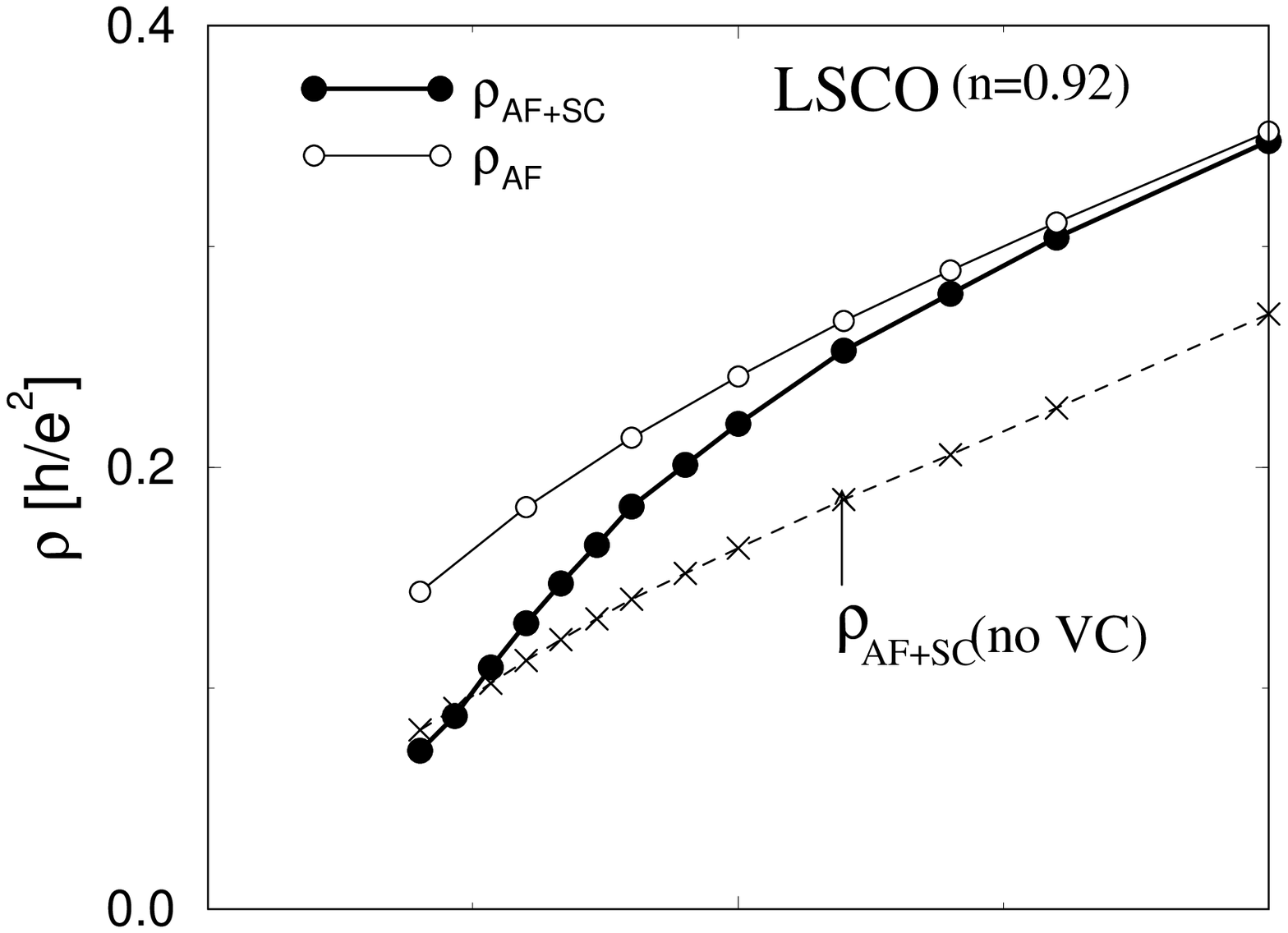,width=5.7cm}
\epsfig{file=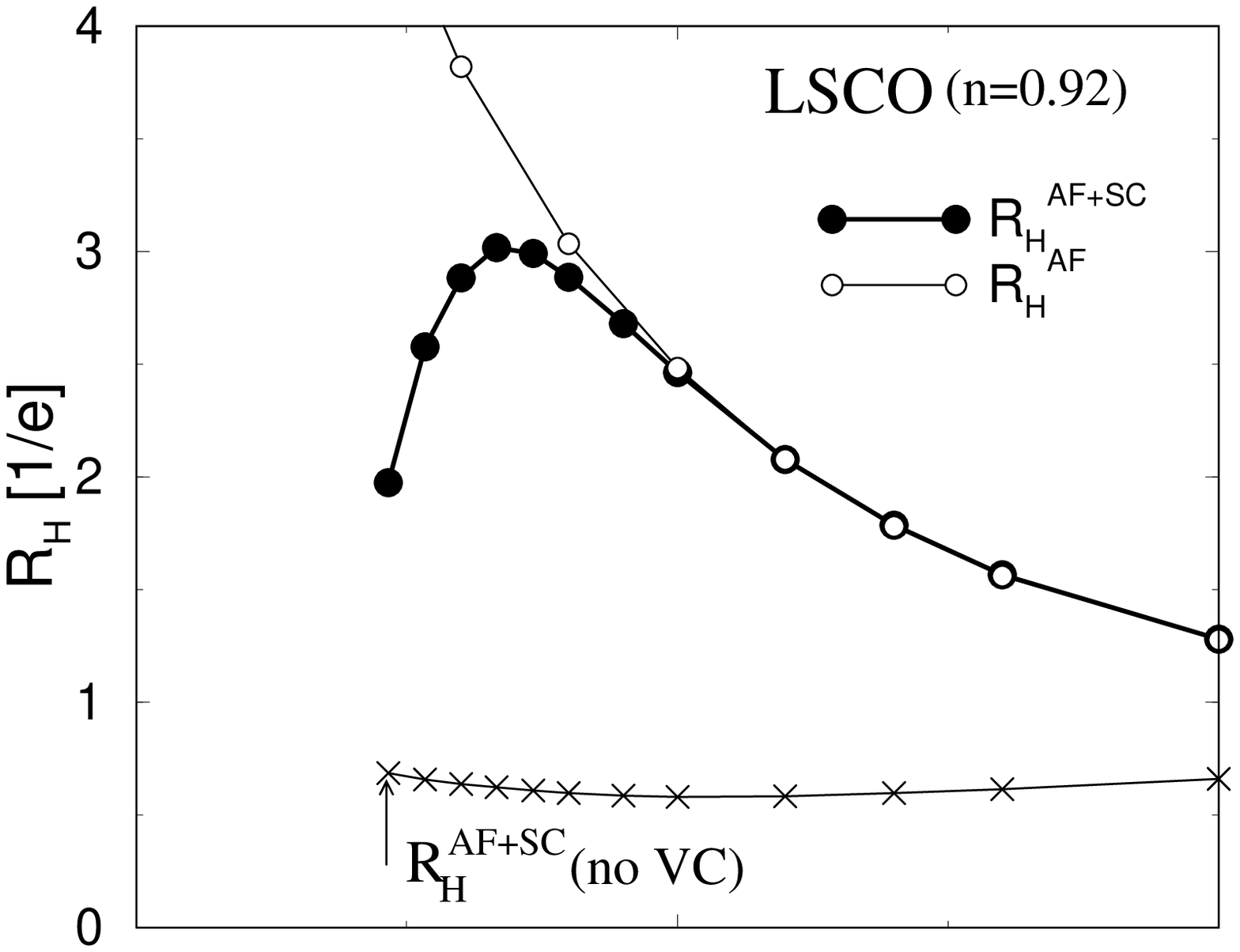,width=5.7cm}
\epsfig{file=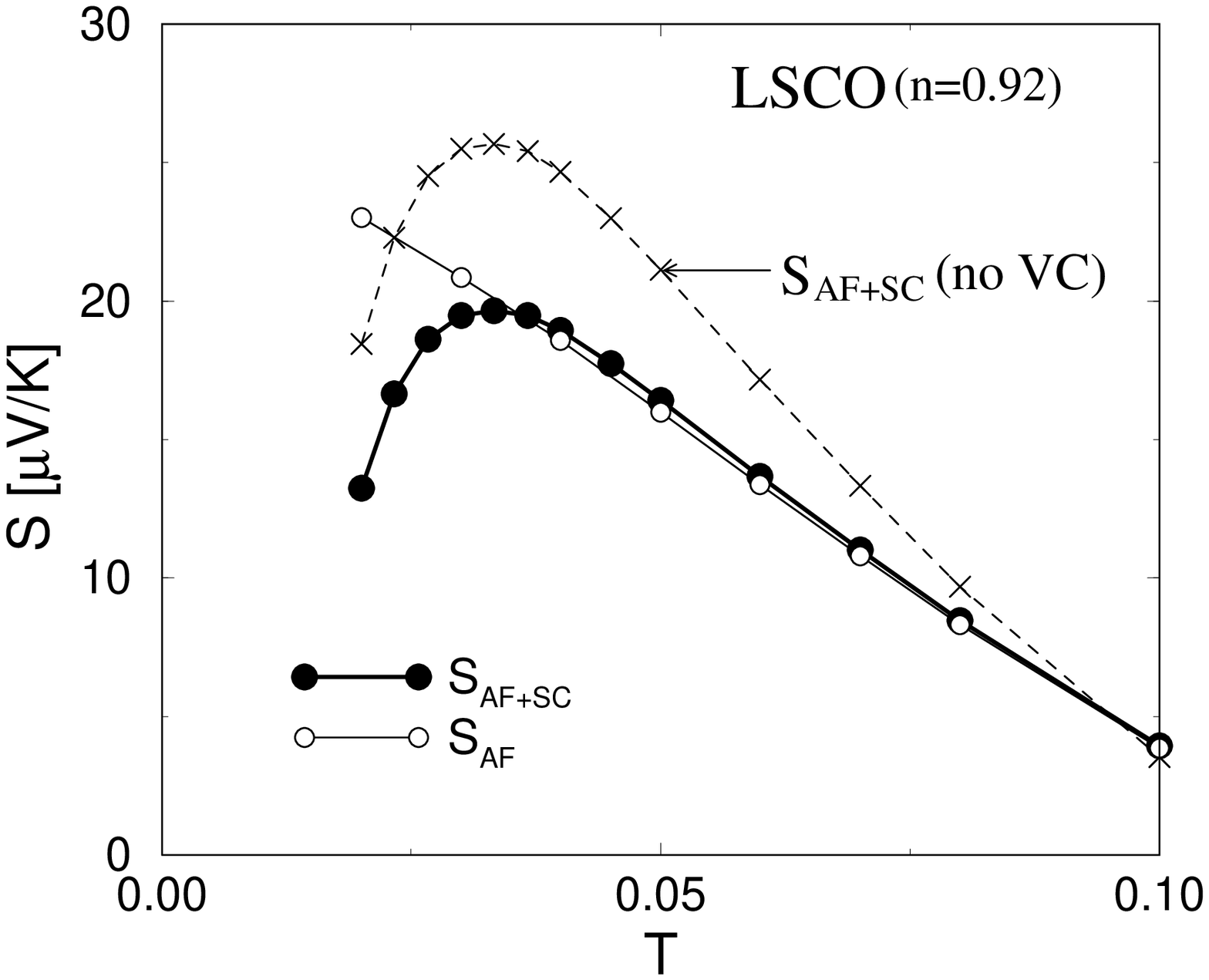,width=5.7cm}
\end{center}
\caption{
Transport coefficients per one layer
studied by the FLEX+T-matrix approximation with VC's (AF+SC).
$h/e^2 = 1.3\times10^6\Omega$.
}
  \label{fig:S}
\end{figure}

Figure \ref{fig:J} show the numerical solution for   
the Bethe-Sapleter equation (\ref{eqn:J-SF-SC}) 
for various $\Delta_{pg}$ at $T=0.02$
 \cite{Hall,Yanase}. 
As $\Delta_{\rm pq}$ increases,  
$|{\vec J}_\k|$ is prominently enhanced  
except for the cold spot (C),  
whereas $\theta_\k^J=\tan^{-1}(J_{\k x}/J_{\k y})$   
is affected only slightly.   
The result is consistent with eq.(\ref{eqn:J-ap})  
because   
$\gamma_\k/\gamma_\k^{\rm FLEX} \approx 1+b\psi_\k^2$  
($b>0$).
As a result,  
owing to the MT-terms by the $d_{x^2-y^2}$-SC fluctuations,  
$|J_\k|$ is enhanced if $\Delta_{\rm pq}$ is large, 
except for the cold spot where $\psi_\k=0$.
A more detailed discussion will be given in 
a future publication
 \cite{future}.
In contrast, we can show that
${\vec Q}_\k(\e) \sim {\vec q}_\k(\e)$
because the effect of the VC for the heat current,
which is not conserved by the electron-electron N-processes,
is small in general
 \cite{Nernst-formula,future}.
This means that ${\vec Q}_\k$ is no more parallel to ${\vec J}_\k$
in the presence of strong AF fluctuations.
This fact is important for understanding the Nernst effect
as will be explained below.

Now we calculate transport coefficients
by using the self-consistent solutions 
of eqs.(\ref{eqn:Q}) and (\ref{eqn:J}).
Figure \ref{fig:S} shows the resistivity, 
the Hall coefficient and the TEP
obtained by the FLEX+T-matrix approximation,
$\rho_{\rm AF+SC}$, $R_{\rm H}^{\rm AF+SC}$ and $S_{\rm AF+SC}$,
for the filling $n=0.92$.
For reference,
$\rho^{\rm AF}$, $R_{\rm H}^{\rm AF}$ and $S_{\rm AF}$
are given by the FLEX approximation.
We see that both $R_{\rm H}^{\rm AF+SC}$ and $S_{\rm AF+SC}$
start to decrease below the pseudo-pap temperature
$T^\ast \sim0.03$, which is consistent with experiments
 \cite{Hall-exp,S-exp}.
Whereas
$R_{\rm H}^{\rm AF}$ and $S_{\rm AF}$
increase monotonously as $T$ decreases
because both quantities
are enhanced due to the VC's 
created by the AF fluctuations
  \cite{Hall,S-HTSC}.
The reason why both 
$R_{\rm H}^{\rm AF+SC}$ and $S_{\rm AF+SC}$
decrease below $T^\ast$ is that the AF fluctuations
are suppressed in the pseudo-gap region,
which was at first discussed in refs.
 \cite{Hall} and \cite{S-HTSC},
and shown numerically in ref. \cite{Yanase} for $R_{\rm H}$.

\begin{figure}
\begin{center}
\epsfig{file=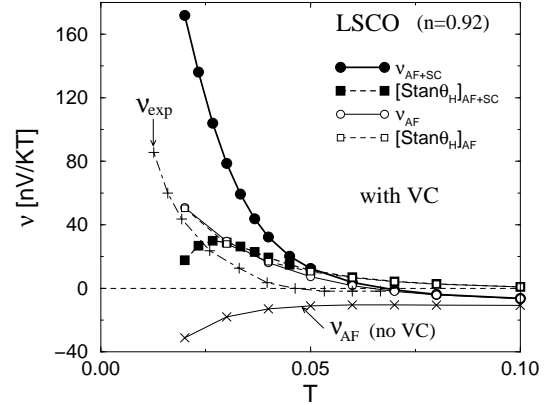,width=7cm}
\end{center}
\caption{
$\nu_{\rm AF+SC}$ and $[S\tan\theta_{\rm H}]_{\rm AF+SC}$
studied by the FLEX+T-matrix approximation with VC's.
$\nu_{\rm exp}$ is the experimental data
reported in ref.[2] for LSCO ($x=0.07$),
assuming that $T=0.1$ corresponds to 300K.
}
  \label{fig:N-LSCO}
\end{figure}

On the contrary,
the drastic enhancement of the Nernst coefficient 
below $T^\ast$ is much mysterious and 
intriguing phenomenon in the pseudo-gap region
 \cite{N-exp,N-exp2}.
Here, we show that
it is naturally explained
as a quasiparticle transport phenomenon,
without assuming thermally excited vortices.
Based on the AF+SC fluctuation theory,
we calculate $\a_{xy}$ given by eq.(\ref{eqn:axy}):
Figure \ref{fig:N-LSCO}
shows the Nernst coefficient 
obtained by the FLEX+T-matrix method, $\nu_{\rm AF+SC}$,
with full MT-type VC's due to AF+SC fluctuations.
We see that $\nu_{\rm AF+SC}$ starts to increase
below $T^\ast$, and its magnitude is consistent 
with experimental values.
In contrast, $S\tan\theta_{\rm H}$ decreases at lower
temperatures, reflecting the suppression of the AF fluctuations.

Here we discuss the reason why
$\nu$ is enhanced in the presence of AF and $d$-SC fluctuations:
The factor $\gamma_\k A_\k$ in eq.(\ref{eqn:axy}) is
rewritten as
\begin{eqnarray}
{\vec Q}_\k \cdot {\vec J}_\k 
 \frac{\d \theta_\k^J}{\d k_\parallel}
+  \left({\vec Q}_\k \times {\vec J}_\k\right)_z
 \frac{\d}{\d k_\parallel} 
 \log \left( {|{\vec J}_\k|}/{\gamma_k} \right) 
 \label{eqn:gA} .
\end{eqnarray}
According to Fig. \ref{fig:J} (ii),
the second term with the factor 
$\frac{\d}{\d k_\parallel}|{\vec J}_\k|$
should cause the enhancement of $\nu$
in the pseudo-gap region:
It does survive because
${\vec J}_\k$ is not parallel to ${\vec Q}_\k$
due to the VC's by strong AF fluctuations.
In contrast, 
this term vanishes identically within the RTA
because of ${\vec v}_\k \parallel {\vec q}_\k$.

\begin{figure}
\begin{center}
\epsfig{file=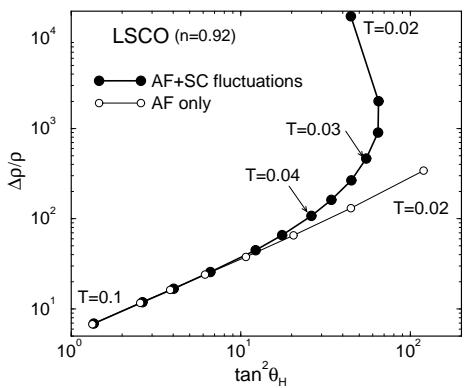,width=7cm}
\end{center}
\caption{
$\Delta\rho/\rho$ by the FLEX+T-matrix approximation
increases drastically in the pseudo-gap region.
}
  \label{fig:MR}
\end{figure}

Finally,
we discuss the MR, $\Delta\rho/\rho$,
in the pseudo-gap region.
Figure \ref{fig:MR}
shows the calculated $\Delta\rho/\rho$
vs $\tan^2\theta_{\rm H}= (\s_{xy}/\s)^2$
for the filling $n=0.92$.
First,
in the FLEX approximation (AF only),
$\Delta\rho/\rho \propto \tan^m\theta_{\rm H}$
and $m\approx 1.5$ is approximately satisfied.
Note that $m\approx 2$ is obtained for $n\sim0.85$
by the FLEX approximation
 \cite{MR-HTSC}, 
which is called the modified Kohler rule
observed in high-Tc cuprates for $T>T^\ast$
 \cite{MR-exp,MR-exp2,Ando-MR}.
In contrast,
according to the present FLEX+T-matrix approximation
where the effect of SC fluctuations is involved (AF+SC),
the relation $\Delta\rho/\rho \propto \tan^m\theta_{\rm H}$
is prominently violated in the pseudo-gap region:
As shown in fig. \ref{fig:MR},
$\Delta\rho/\rho$ starts to increase abruptly
below $T^\ast\sim 0.03$,
while the suppression of $\tan\theta_{\rm H}$ 
starts at the same time.
The obtained result is very well consistent with experiments
 \cite{MR-exp,MR-exp2,Ando-MR}.

Now, we discuss why
$\Delta\rho/\rho$ increases drastically below $T^\ast$:
The expression for magnetoconductance
($\Delta\s_{xx}$), which is given by eq.(7) of ref.
 \cite{MR-HTSC},
has a term proportional to 
$( \ \frac{\d}{\d k_\parallel}|J_\k| \ )^2$.
As discussed before,
this factor causes the drastic increase of 
$\Delta\rho/\rho$ below $T^\ast$
(see fig. \ref{fig:J} (ii)).
As a result, 
we find the reason why both $\nu$ and $\Delta\rho/\rho$
are enhanced in the pseudo-gap region.
In the meanwhile,
other quantities like $\rho$, $R_{\rm H}$
and $S$ decrease below $T^\ast$ because of the lack of 
the factor $\frac{\d}{\d k_\parallel}|J_\k|$.

We note that the positive MR below $T^\ast$
is frequently ascribed to the magnetic field suppression of
the transport due to SC fluctuations, 
which is theoretically expressed 
as the Aslamazov-Larkin (AL) term.
However, one have to assume a
rather large coherent length 
in the under-doped region 
($\raisebox{-0.75ex}[-1.5ex]{$\;\stackrel{<}{\sim}\;$}$40\r{A})
to fit the observed huge MR
 \cite{Ando-MR}. 
Instead, the present work naturally explains
the increase of $\Delta\rho/\rho$ below $T^\ast$, 
as well as $\nu$, in terms of the quasiparticle transport phenomena, 
which is expressed as the infinite series of MT terms.

In summary, 
we studied the origin of the transport anomaly
in the pseudo-gap region
using the FLEX+T-matrix approximation.
Below $T^\ast$ in hole-doped compounds,
the resistivity, the TEP and the Hall coefficient
decrease moderately,
whereas the Nernst coefficient and the
MR increase drastically.
In the present study,
we could reproduce the characteristic behaviors
of these coefficients satisfactorily.
Especially, the drastic increase of $\nu$ and $\Delta\rho/\rho$
below $T^\ast$ 
is naturally explained as the quasiparticle origin,
by taking the VC's due to the AF and SC fluctuations correctly.
Thus, unusual transport properties of high-Tc cuprates are 
well explained as the quasiparticle transport phenomena.

We are grateful to 
H. Eisaki, Y. Ando, K. Yamada, T. Saso, K.Ueda and R. Ikeda
for comments and discussions.

\vspace{-6mm}


\end{multicols}


\begin{thebibliography}{99}

\vspace{-16mm}

\bibitem{rho-exp}
 e.g., K. Takenaka et al.: Phys. Rev. {\bf B50} (1994) 6534.

\bibitem{Hall-exp}
 e.g.,Z.A. Xu et al.: cond-mat/9903123.

\bibitem{MR-exp}
A. Malinowski et al.: cond-mat/0108360.

\bibitem{MR-exp2}
Y. Ando et al.: Phys. Rev. {\bf B60} (1999) R6991.

\bibitem{Ando-MR}
 Y. Ando et al.: Phys. Rev. Lett. {\bf 88} (2002) 167005. 

\bibitem{S-exp}
 e.g.,  
J. Takeda et al:
Physica C {\bf 231} (1994) 293.

\bibitem{N-exp}
Z.A. Xu et al:
 Nature {\bf 406} (2000) 486.  

\bibitem{N-exp2}
C.C. Capan et al.: cond-mat/0108277.

\bibitem{Baym}
 G. Baym and L.P. Kadanoff: Phys. Rev. {\bf 124} (1961) 287.

\bibitem{Hall} 
H. Kontani et al.:
 Phys. Rev. B {\bf 59} (1999) 14723.

\bibitem{MR}
 H. Kontani: Phys. Rev. B {\bf 64} (2001) 054413.

\bibitem{MR-HTSC}
 H. Kontani: J. Phys. Soc. Jpn. {\bf 70} (2001) 1873.

\bibitem{S-HTSC}
 H. Kontani: J. Phys. Soc. Jpn. {\bf 70} (2001) 2840.

\bibitem{SCR}
 T. Moriya and K. Ueda: Adv. Physics {\bf 49} (2000) 555.

\bibitem{Bickers}  
 N.E. Bickers et al.: 
Phys. Rev. B {\bf 43} (1991) 8044

\bibitem{Levin}
 Q. Chen, I. Kosztin, B. Jank{\' o} and K. Levin:
 Phys. Rev. Lett {\bf 81} (1998) 4708,
 and references are therein.

\bibitem{Koikegami}
 Koikegami and K. Yamade: Soc. Jpn. {\bf 69} (2000) 768.

\bibitem{Yanase}
 Y. Yanase: J. Phys. Soc. Jpn. {\bf 71} (2002) 278,
 and references are therein.

\bibitem{Kobayashi}
 A. Kobayashi, A. Tsuruta, T. Matsuura and Y. Kuroda:
 Soc. Jpn. {\bf 69} (2000) 225,
 and references are therein.

 \bibitem{Mahan}
 M. Jonson et al.:
Phys. Rev. B {\bf 42} (1990) 9350.

\bibitem{Kohno}
 H. Kohno et al.:
Prog. Theor. Phys. {\bf 80} (1988) 623.

\bibitem{Nernst-formula}
 H. Kontani : cond-mat/0206501.

\bibitem{Eliashberg}
 G.~M. Eliashberg :
 Sov. Phys. JETP {\bf 14} (1962) 886.

\bibitem{future}
 H. Kontani : in preparation.

\end{thebibliography}
\end{document}